\documentclass[a4paper,twoside]{article}
\newcommand{\affil}[1]{$^{\rm #1}$}
\usepackage[authoryear]{natbib}
\usepackage{multirow}
\bibpunct{(}{)}{;}{a}{}{,}
\usepackage{graphicx}
\usepackage{url}
\usepackage{color}
\usepackage{subcaption}
\usepackage[T1]{fontenc}
\date{} 
\graphicspath{ {images/} }
\title{\large\bf\flushleft Collaborative workspaces to accelerate discovery}
\author{\parbox{\textwidth}{\flushleft
\vspace{-0.5cm}
{\it Bernard Meade\affil{A,B,H}, Christopher Fluke\affil{A}, Jeff Cooke\affil{A}, Igor Andreoni\affil{A,D,G}, Tyler Pritchard\affil{A}, Christopher Curtin\affil{A}, Stephanie R. Bernard\affil{C}, Albany Asher\affil{A}, Katherine J. Mack\affil{C,D,E}, Michael T. Murphy\affil{A}, Dany Vohl\affil{A}, Alex Codoreanu\affil{A,D}, Sr{\dj}an M. Kotu\v{s}\affil{A}, Fanuel Rumokoy\affil{C}, Chuck Horst\affil{F} and Tristan Reynolds\affil{C}}\\
\vspace{0.4cm}
{\small \affil{A}\,Centre for Astrophysics and Supercomputing, Swinburne University of Technology, PO Box 218, Hawthorn, Australia, 3122}\\
{\small \affil{B}\,Research Platform Services (Doug McDonell Building), The University of Melbourne, Victoria 3010, Australia}\\
{\small \affil{C}\,School of Physics (David Caro Building), The University of Melbourne, Victoria 3010, Australia}\\
{\small \affil{D}\,ARC Centre of Excellence for All-sky Astrophysics (CAASTRO), The University of Sydney, NSW 2006, Australia}\\
{\small \affil{E}\,ARC Centre of Excellence for Particle Physics at Terascale (CoEPP), School of Physics, The University of Melbourne, Victoria 3010, Australia}\\
{\small \affil{F}\,Department of Astronomy, San Diego State University, San Diego, CA 92128-1221, United States}\\
{\small \affil{G}\,Australian Astronomical Observatory, PO Box 915, North Ryde, NSW 1670, Australia}\\
{\small \affil{H}\,Email: bmeade@unimelb.edu.au}}}
\begin{document}
\twocolumn[
\begin{changemargin}{.8cm}{.5cm}
\begin{minipage}{.9\textwidth}
\vspace{-1cm}
\maketitle
%
%
\small{\bf Abstract:} 


By applying a display ecology to the {\em Deeper, Wider, Faster} proactive, simultaneous telescope observing campaign, we have shown a dramatic reduction in the time taken to inspect DECam CCD images for potential transient candidates and to produce time-critical triggers to standby telescopes.  We also show how facilitating rapid corroboration of potential candidates and the exclusion of non-candidates improves the accuracy of detection; and establish that a practical and enjoyable workspace can improve the experience of an otherwise taxing task for astronomers.  We provide a critical road-test of two advanced displays in a research context -- a rare opportunity to demonstrate how they can be used rather than simply discuss how they might be used to accelerate discovery.


\medskip
\medskip
\end{minipage}
\end{changemargin}
]
\small

\section{Introdution}
\label{sct:introduction}

Investment from research institutions and governments in new astronomical facilities, scientific instruments, and high-performance computing capabilities occurs at a great scale. The shared resources are typically heavily oversubscribed and time allocation on these instruments is extremely competitive.  Simultaneously, new science, such as the search for fast transients, places an even bigger strain on the available resources as it requires several telescopes for coordinated observation, with additional telescopes to be on standby for immediate re-pointing if a significant event occurs.  Therefore, it is imperative that all aspects of the scientific workflows engendered by this research infrastructure are scrutinised.  While much effort is expended evaluating the processes for the observing, computation and storage components of a workflow, only recently has attention been given to the operational workspace in which humans interact with the technological systems.  For many years, the standard computer display served to present all manner of digital content, from text to graphs to images, with little consideration as to the appropriateness of the display to the content.  In order to accelerate discovery, workspaces that facilitate collaboration and understanding in real-time, both in situ and remotely, are fast becoming essential.  A carefully considered {\em display ecology} \citep{huang2006,chung2015} that addresses specific visualisation tasks are a key component to achieving satisfactory scientific outcomes.

\subsection{The role of the display}
\label{sct:roleofdisplay}
Computer displays, or monitors, have become such an integral component of the astronomer's scientific toolset, that it can be easy to overlook their significance or impact on the research workflow.   It can be tempting to continue to use a display -- even when its size or resolution begin to limit productivity -- simply because it is available or on-hand rather than assessing the capabilities of an alternative solution.

It may be necessary to increase the physical size of the display in order to: inspect very large datasets (e.g. thousands of DECam{\footnote{\url{http://www.ctio.noao.edu/noao/node/1033}}} 520 megapixel mosaics), where each image is orders of magnitude larger than the display it is being viewed on; achieve meaningful collaboration; and make discoveries that require the rapid or simultaneous validation additional astronomers and experts. This can be achieved through the purchase of a bigger -- and thus usually more expensive -- monitor, using a digital data projector, or by adding additional screens to the desktop.  Occasionally, it requires a more radical re-thinking of what a display can be.

Some of the earliest work on alternative displays for astronomy was by \citet{fomalont1982} and \citet{rots1986}.  \citet{norris1994} examined the potential role for qualitative, comparative and quantitative visualisation, and \citet{fluke2006} presented options for collaborative environments: multi-projector tiled displays; digital domes; and large-scale collaborative, stereoscopic exploration of three-dimensional datasets, {\em viz} The Virtual Room.

\subsection{Collaboration}
\label{sct:collaboration}
Scaling up a display (in terms of physical size and the number of pixels) necessitates a move away from the desktop.  There is an  opening up of space around (or in front of) the display, encouraging researchers to stand up, move around and share the workspace with their colleagues.  These are key elements that can turn visualisation and inspection of data from a solitary pursuit into a collaborative one.


The value of collaboration in scientific discovery has been at the heart of endeavours such as the CAVE and the OptIPortal projects \citep{cruz1992, smarr2003, smarr2005, defanti2009, defanti2010, febretti2013}.  Placing multiple researchers in the same physical space and allowing them to interact with data together allows them to experience a shared engagement with the information. Contrast this with coincident engagement when they experience the data at separate locations at the same time, even while in contact via communication technologies.

Advanced collaborative workspaces with large, immersive display technologies have been in use around the world for over a decade, yet their impact on the research landscape has been relatively limited.  While these facilities have been used as educational tools and high-impact demonstration environments \citep{adlerweb2007, sdoweb2015, qutcubeweb2016}, there is a dearth of published research that identifies dedicated collaborative display environments as a critical component in the workflow that has produced new scientific outcomes. Furthermore, beyond the advantages of the technology itself, the value of bringing astronomers together in a single space to collaborate in real-time is considerable.

\subsection{Tiled Display Walls}
\label{sct:tdw}
On the face of it, it seems likely advanced displays should lead to more rapid scientific discovery and would therefore be deemed essential.  A specialised tool that improves engagement, by enhancing immersion or providing access to many more pixels, should afford greater insight to its users and scientific outcomes should follow.  In reality, that has not been the case.  But perhaps the issue is not that the displays themselves are not capable of achieving such goals, but that they have yet to find the right place in the research workflow. 

\citet{meade2014} consider some of these possibilities, while also testing the assumption that a display environment such as a Tiled Display Wall (TDW) can actually improve research outcomes.  

Clustering homogeneous computer displays to simulate a single continuous display canvas has been possible for some time.  In this approach, the physical displays are placed in a fixed array and connected to one or more computers, often referred to as nodes or workers.  These nodes are coordinated by a single head node, which typically does not take part in the display environment.  

The content being displayed on any individual screen is synchronised by the head node to provide the user with the appearance of a single display.  In this way, media content can be moved around the entire display almost seamlessly.  High resolution images (or movies) that greatly exceed the resolution of an individual display can then be shown at full resolution across several displays.  The only interruption to the display space are the screen mountings, called bezels, at the edge of each screen.

The \citet{meade2014} study concluded that in certain circumstances, and for certain people, a TDW will improve a user's ability to find small targets (185 $\times$ 145 pixels) within a much larger image (12,000 $\times$ 5812 pixels).  It also highlighted a tendency for individuals to prefer physical navigation, that is, the use of physical body movements such as eyes, head and the whole body, to virtual navigation, which uses computing interfaces such as keyboard and mouse, when inspecting very large images.  These findings were consistent with the outcomes of \citet{ball2005a, ball2005b} and \citet{andrews2010, andrews2011}.

\subsection{The display ecology}
\label{sct:displayecology}

\begin{figure*}
\begin{center} 
\includegraphics[width=16cm, angle=0]{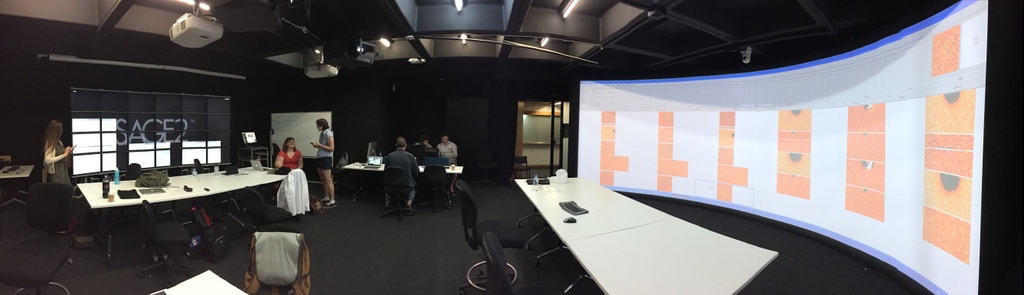}
\caption{A panoramic view of the workspace used for O1, showing the TDW at the left of the image, the review and control stations in the middle and the curved projection screen to the right. The whiteboard shown centre left was used to log potential candidates for review, as well as other important details including telescope on sky times.}
\label{fig:wholeroom}
\end{center}
\end{figure*}

By construction, \citet{meade2014} used an artificial context that {\em resembled} a research activity from astronomy: visual inspection of an image looking for known types of objects in unknown locations. However, it also identified another context in which a TDW might be useful: parallel inspection of many images, as opposed to a single extremely large image.   Yet a TDW is not suitable for all types of data, and may enforce technological limitations of its own (such as the limited availability or functionality of software that can drive the display -- see section \ref{sct:probstdw}). 

Combining display technologies to form a display ecology \citep{huang2006, chung2015} offers a best of all worlds approach.  While each display can overcome a particular hurdle to understanding, they can also ignore or exacerbate others.  Using the right display for the right content in concert improves a researcher's ability to draw on many sources to construct a more complete mental picture of the science at hand.

\subsection{Overview}
In this paper, we present a case study based on our use of a collaborative workspace to support the {\em Deeper, Wider, Faster} initiative.    {\em Deeper, Wider, Faster} is a coordinated, contemporaneous, multi-wavelength observing program.  It aims to make rapid, real-time identification of fast transients, i.e. those with a duration of milliseconds to hours, including Fast Radio Bursts (FRBs), Gamma-ray bursts, flare stars, {kilonov\ae} and supernova shock breakouts, using telescopes across the globe and in orbit.  

A full description of the observing strategy, discovery pipeline and detections from the first four campaigns are outside of the scope of this paper.  Full details of the {\em Deeper, Wider, Faster} program may be found in Cooke et al. ({\em in prep}), Andreoni et al. ({\em submitted}) and future papers.   We discuss only those aspects of observation, discovery and analysis that informed our approach to understanding, adopting and improving the display ecology.

The {\em Deeper, Wider, Faster} pilot program (see Section \ref{sct:dwf}) raised a number of issues relating to large-format image inspection and collaboration.   While preparing for future campaigns, a TDW was identified as being a strong candidate to eliminate several key problems with the existing desktop-based workflow.  This visualisation environment was augmented by the use of large-format curved display, with the two displays working in tandem.

In the remainder of this paper, we present the collaborative workspace used for the {\em Deeper, Wider, Faster} 17-22 December 2015 UT (operational run 1: O1) and the 26 July - 7 August 2016 UT (operational run 2: O2) campaigns.  During these observing runs, up to 12 astronomers at a time shared the workspace. 

For O1, our solution used two advanced displays, a 98 Megapixel tiled display wall and a curved projection screen (6.9m circular segment with 2.56m radius, 2.2m height), co-located in the Advanced Immersive Environment at the University of Melbourne -- see Figure \ref{fig:wholeroom} -- along with several laptops and desktop computers.  

One of the main objectives of the display ecology is to enable rapid identification (in minutes) of fast-evolving transient events to inform other telescopes of the location of the discoveries and to "trigger" them to rapidly move to obtain spectroscopy or additional imaging of the objects before they fade.

During O1, several candidates were identified as potential spectroscopic trigger candidates, and a number of triggers were sent.  For example, a live trigger was sent to Gemini-South for spectroscopic follow up and did result in the successful acquisition of the spectrum of an extragalactic transient currently under investigation.  Moreover, the process proved valuable in uncovering CCD artefacts, amplifier crosstalk and other effects that can produce ``fake'' fast transients.  Also, the observing team were able to critically road-test how advanced displays can be used -- a rare alternative to previous discussions of how they might be used \citep{fomalont1982, rots1986, norris1994, fluke2006}.  

\begin{figure*}
\begin{center} 
\includegraphics[width=16cm, angle=0]{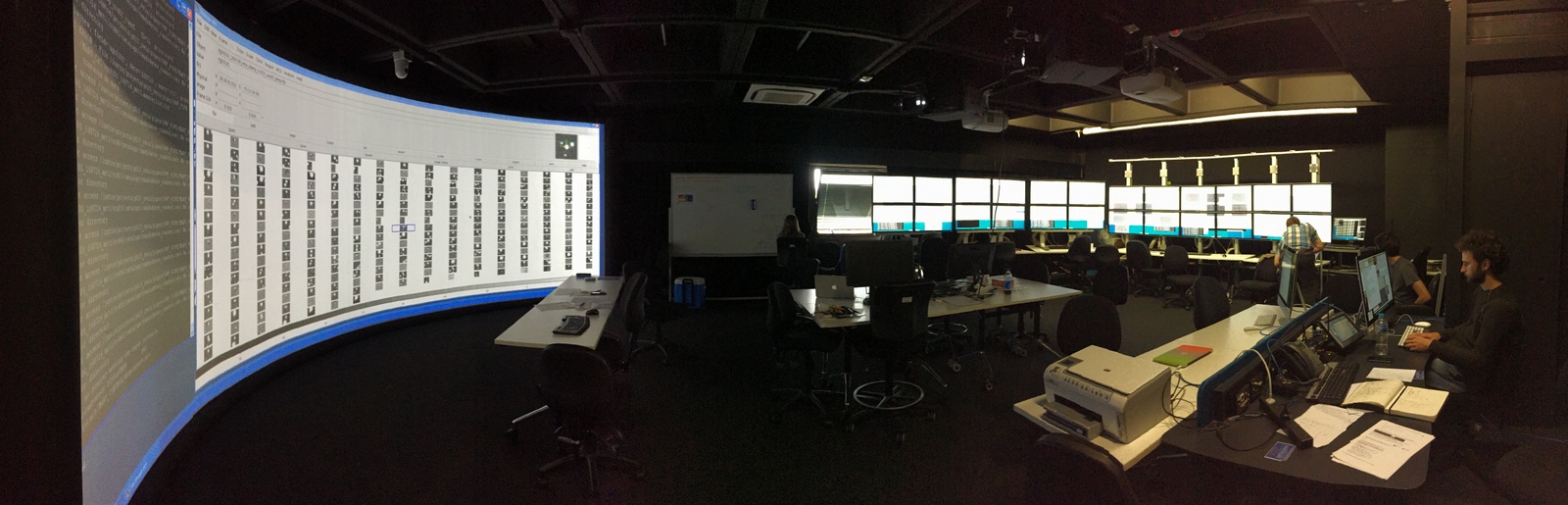}
\caption{A panoramic view of the updated workspace for O2, showing the reconfigured TDW at the right of the image, the review and control stations in the middle and the curved projection screen to the left. }
\label{fig:wholeroom2}
\end{center}
\end{figure*}

For O2, the processing workflow underwent several additional improvements based on our insights, and user feedback, from O1. Specifically, we rearranged the Advanced Immersive Environment at the University of Melbourne -- see Figure \ref{fig:wholeroom2}, and integrated additional online tracking of candidates.    The improved workflow, including advancements in the automatic candidate detection pipeline, resulted in three triggers sent to the Gemini-South Observatory and four triggers to the South African Large Telescope (SALT).  In addition, round 570 spectra were obtained via the Australian Astronomical Telescope (AAT) with over 50 targets identified for follow-up with the Zadko Telescope (University of Western Australia) and Skymapper (Australian National University).  In all, tens of thousands of candidate variable and transient objects were detected during this run.

Through a combination of pre-campaign questions, observations of usage patterns during the observing period, and post-campaign reflection we:

\begin{enumerate}
\item Demonstrate that careful design of a collaborative workspace can greatly improve the rate at which CCD images can be inspected;
\item Show how facilitating rapid corroboration of potential candidates and the exclusion of non-candidates improves the accuracy of detection; and
\item Establish that a practical and enjoyable workspace can improve the experience of an otherwise taxing task for astronomers.
\end{enumerate}

The paper is set out as follows. In Section \ref{sct:dwf} we describe the pilot program for the {\em Deeper, Wider, Faster} observing campaign, and identify the visualisation-based bottlenecks inherent in the original workflow.  In Section \ref{sct:run3} we discuss the setup for the December 2015 (O1) and in Section \ref{sct:results} we evaluate the collaborative workspace and the impact of the display technology on the workflow.  The changes implemented for July/August 2016 (O2) are described in Section \ref{sct:run4}. In Section \ref{sct:discussion} we discuss planned improvements to both the process and the technological workflow in order to improve future scientific outcomes.  Concluding remarks are made in Section \ref{sct:conclusion}.

\section{{{\em Deeper, Wider, Faster}}}
\label{sct:dwf}



{\em Deeper, Wider, Faster} primarily uses the Dark Energy Camera (DECam; \citep{diehl2012, flaugher2012}) on the Blanco 4-m Telescope at the Cerro Tololo Inter-American Observatory (CTIO) in Chile to observe a region of the sky. These fields are simultaneously observed at radio wavelengths by the Parkes radio telescope and the upgraded Molonglo Observatory Synthesis Telescope (MOST) in Australia, the NASA {\em Swift} Space Telescope in low-Earth orbit, and occasionally other facilities such as the Very Large Array (VLA) in the US.  

Should a suitable transient candidate be identified, such as a potential optical counterpart to an FRB, alerts were to be sent to additional telescopes, such as the Gemini-South Observatory, for targeted optical/infrared spectroscopic follow-up. In order to confirm an event, obtain its redshift, localise it, obtain its host galaxy properties, and understand its nature, spectroscopy needs to be acquired within minutes of the detection before the fast transient event fades, thus the urgency to process and identify candidate sources.  

\begin{figure*}
\begin{center} 
\includegraphics[width=16cm, angle=0]{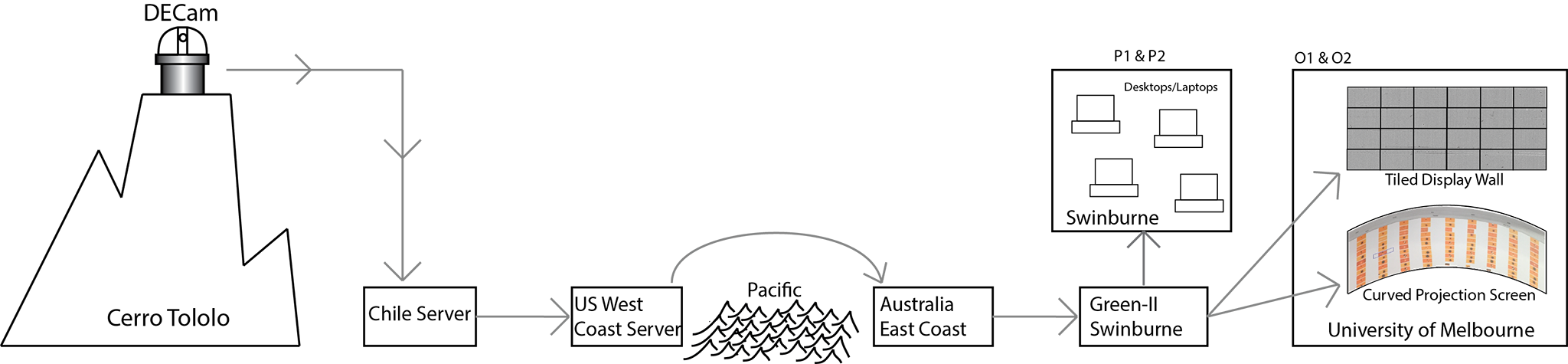}
\caption{While data from many telescopes was collected, the focus of the data inspection optimisation for O1 and O2 was on the optical image data captured with the DECam imager in Chile that was then transferred to Green II supercomputer at Swinburne University for processing.  In the pilot programs, P1 and P2, the images were inspected on desktop and laptop computers in Chile and Swinburne University.  In O1, after processing, the images were transferred to the University of Melbourne for inspection on the Tiled Display Wall and on the curved projection screen (see Section \ref{sct:introduction}). In O2, the images were inspected on Tiled Display Wall reconfigured as 6 individual workstations (see Section \ref{sct:run4}), and on the curved projection screen. }
\label{fig:dwftransfer}
\end{center}
\end{figure*}

In the example of FRBs, optical and spectroscopic counterparts have yet to be identified and their behaviour at wavelengths other than radio are completely unknown, making it challenging to design and use a purely automatic detection pipeline to identify possible progenitors.  For now, there is an important role for visual inspection of images and potential candidates at all stages of the workflow.  This includes making judgements as to the likelihood that a potential candidate could be a counterpart, performing quality control tasks, or making serendipitous discoveries of as yet undetermined transient objects.

\subsection{The pilot program}
\label{sct:phase1}

The initial {\em Deeper, Wider, Faster} observing campaigns were held from 13-16 January 2015 UT (pilot run 1: P1) and 27-28 February 2015 UT (pilot run 2: P2).  Figure \ref{fig:dwftransfer} shows the flow of data from the DECam imager on the Blanco telescope at Cerro Tololo in the Chilean Andes to Swinburne University of Technology, Melbourne, Australia.  While several other telescopes were involved, this paper focuses on the collaborative workspace used for visualisation and review of DECam images to discover transient sources.

For P1, several members of the observing team were situated at Cerro Tololo, to view and interact with the DECam images directly prior to transfer to Australia.  The observers were able to use a 6-panel tiled display (consisting of 27" desktop LCD monitors @ 1920$\times$1200 pixels on a standard desktop computer), however individual CCD images could not be expanded across the full display. Instead, the screens were used to display 6 concurrent CCD images on individual monitors along with researchers' desktops and laptops.    For P2, most of the team were located at Swinburne where all the analysis was performed using only desktop or laptop computers.

The pilot runs P1 and P2 were designed as an opportunity to uncover and deal with obstacles that typically prevent real-time fast transient detection and study.  This process identified the need for sophisticated visual inspection and in turn, the development of a supporting display ecology.  In addition, P1 and P2 brought about the development of software to provide real-time data compression, processing, and analysis, and the software for real-time candidate identification.

The DECam CCD images were subtracted from a template to produce difference images.  This provides the best opportunity to identify significant changes in the images since the template was captured that might indicate an event of interest.  While perfect alignment is not possible and many artefacts remain after the subtraction, the combination of automatic catalogue lookup and the eyes of experienced astronomers are able to find the objects of interest. At 4096 $\times$ 2048 pixels, these were significantly larger than the resolution of the standard displays used (up to 2560 $\times$ 1440 pixels), therefore the inspection relied heavily on {\em virtual navigation} -- zooming and panning -- to search for potential candidates.  This process was performed in parallel for each of the DECam CCDs, and occasionally with sections of the full DECam mosaic (60 CCD images).  

The full images were examined (1) to understand the context and, equally, (2) to determine if the sources were CCD artefacts such as amplifier crosstalk (which requires full CCD inspection).  
As the physical scale of potential candidates is unknown, there is a risk of overlooking a feature of interest due to pixel subsampling.  On a display that is considerably smaller than the image size, viewing at native resolution requires methodical scanning of the images, which is tedious.  Zooming in and out on features of potential interest makes objects on the edge of detectability very difficult to find.

While the impact of pixel subsampling was not investigated in depth, a qualitative assessment was made by the authors by comparing the native resolution CCD images with a full screen version on the 2560 $\times$ 1600 displays.  At this resolution, the CCD images are being displayed at less than 50\% of their native resolution and the authors found the potential candidate detection to be far more difficult to perform.

The process of inspecting the individual CCD images in this manner was deemed to be a significant bottleneck in the selection of potential candidates for follow-up study.  Even though the process was slow, it was still essential in determining potential candidates before a trigger to engage additional telescopes could be sent.  The process was also greatly complicated when several CCD images needed to be compared.  Finally, the lack of physical space in front of standard desktop displays limited collaboration, forcing the researchers to use multiple independent computers and displays, thus reducing the effectiveness of a parallel search.

With more observing runs planned, it was important to change the processes.   The development of an automatic detection pipeline (described in Andreoni et al. {\em submitted}) and the use of advanced displays were expected to significantly improve productivity.  With an emphasis on decreasing the time spent inspecting each CCD image, eliminating virtual navigation, enhancing collaboration and integrating the automatic search more completely into the visual search, a new collaborative workspace was needed.

\subsection{Workspace requirements}
\label{sct:requirements}


\begin{figure*}
\begin{center} 
\includegraphics[width=15cm, angle=0]{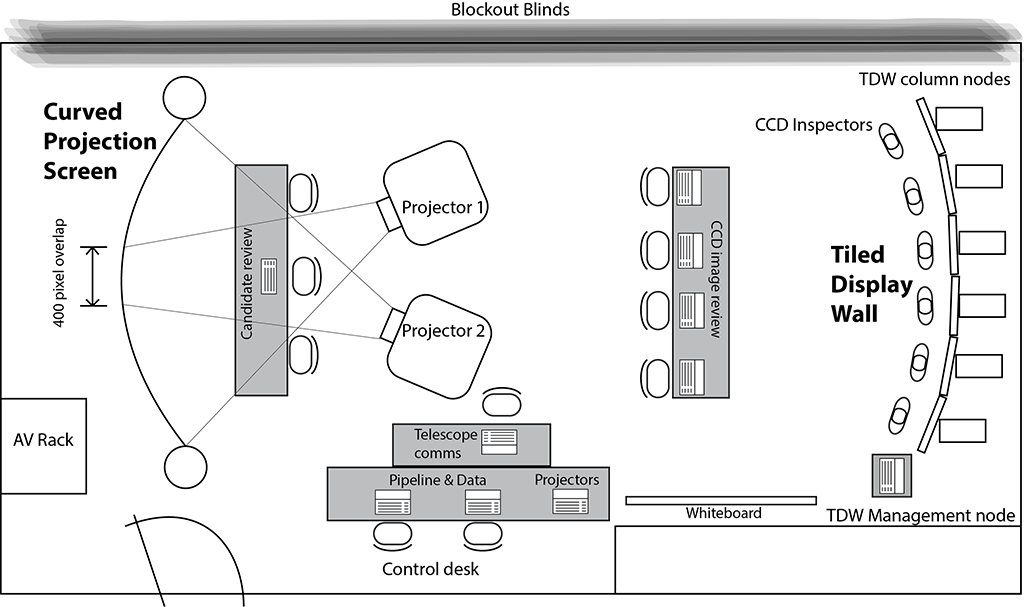}
\caption{Floor plan of the Advanced Immersive Environment at the University of Melbourne for O1.  The room configuration allowed the two principal activities, i.e. reviewing the software identified candidates on the curved screen and inspection of the CCD difference images on the TDW, to be conducted independently while supporting collaboration between these tasks.  The control desk had an excellent view of both sides of the room, and team-members here could easily respond to requests from either side. }
\label{fig:layout}
\end{center}
\end{figure*}

\begin{table*}
\centering
\caption{Hardware specifications of the principal workstations and projectors used during O1}
\label{lbl:hardware}
\begin{tabular}{lll}
\textbf{TDW SAGE2} & \textbf{Specification} & \textbf{Display}    \\ \hline
Head node          & Virtual Server (NeCTAR Research Cloud)16 vCPUs, & No attached display\\
& 64GB RAM, 500GB Volume Storage, 10GB network, \\
  & Ubuntu 14.04LTS  \\
Management node    & Dell T3400, Quad-core Intel, 2.4GHz, 16GB RAM, & 1$\times$19inch display\\
& 500GB HDD, 1GB network, Ubuntu 14.04LTS \\
& (1680$\times$1050)   \\
Display nodes ($\times$6) & Dell T3400, Quad-core Intel, 2.4GHz, 16GB RAM, & 4$\times$Dell Ultrasharp \\
& 500GB HDD, 1GB network, Quadro FX570, &30inch display \\
& 2GB VRAM, Ubuntu 14.04LTS &(2560$\times$1600)  \\ \hline
\textbf{Curved Screen} & \textbf{Specification} \\ \hline
Head node          & Dual core Xeon 3.00GHz, 3GB RAM, 500GB HDD, \\
& 1GB network, Windows XP SP4 \\
Christie Projectors    & 2$\times$1920$\times$1200 (400 pixel overlap for image blending), \\
& fast phosphor, 120Hz for active stereo \\
  &(not used in this experiment) \\ \hline
\textbf{Control station} & \textbf{Specification} \\ \hline
Pipeline \& Data          & iMac 27-inch (diagonal) LED-backlit display \\
&(2560x1440) \\\hline
\end{tabular}
\end{table*}

In preparing for O1 (17-22 December 2015 UT), there was a clear need to improve the visual inspection workflow.  The five key requirements were:
\begin{itemize}
\item Decrease the time to inspect a full CCD, or even the entire 60-CCD field of view of DECam: optimising the time taken to complete the visual inspection of difference images is critical for confirming suitable candidates for rapid follow up, reducing the time from days or hours to minutes. 
\item Remove virtual navigation: by eliminating the need to pan and zoom images, inspectors can be more confident of complete coverage of an image and reduce the risk of overlooking potential candidates.
\item Enhance collaboration: having inspectors working independently but immediately adjacent provides rapid corroboration of potential candidates, with minimal disruption to the inspection workflow.
\item Integrate the automatic search more completely into the visual search: make better use of the astronomers' time to look at the most important things and to enhance the mutually supporting review processes.
\item Completeness test: use the visual inspection as a means to provide a completeness test and training set for the software candidate selection.
\end{itemize}

Several constraints were imposed on the development of the display ecology, including:
\begin{itemize}
\item No funding to secure hardware resources specifically to support the display ecology;
\item No staff resources for developing a bespoke software solution to support the display ecology; and
\item Relocating the compute resources and inspection team to Chile to reduce the impact of the physical separation between the workflow and the capture device, DECam, would have been prohibitively expensive.
\end{itemize}

\begin{figure*}
\begin{center} 
\includegraphics[width=16cm, angle=0]{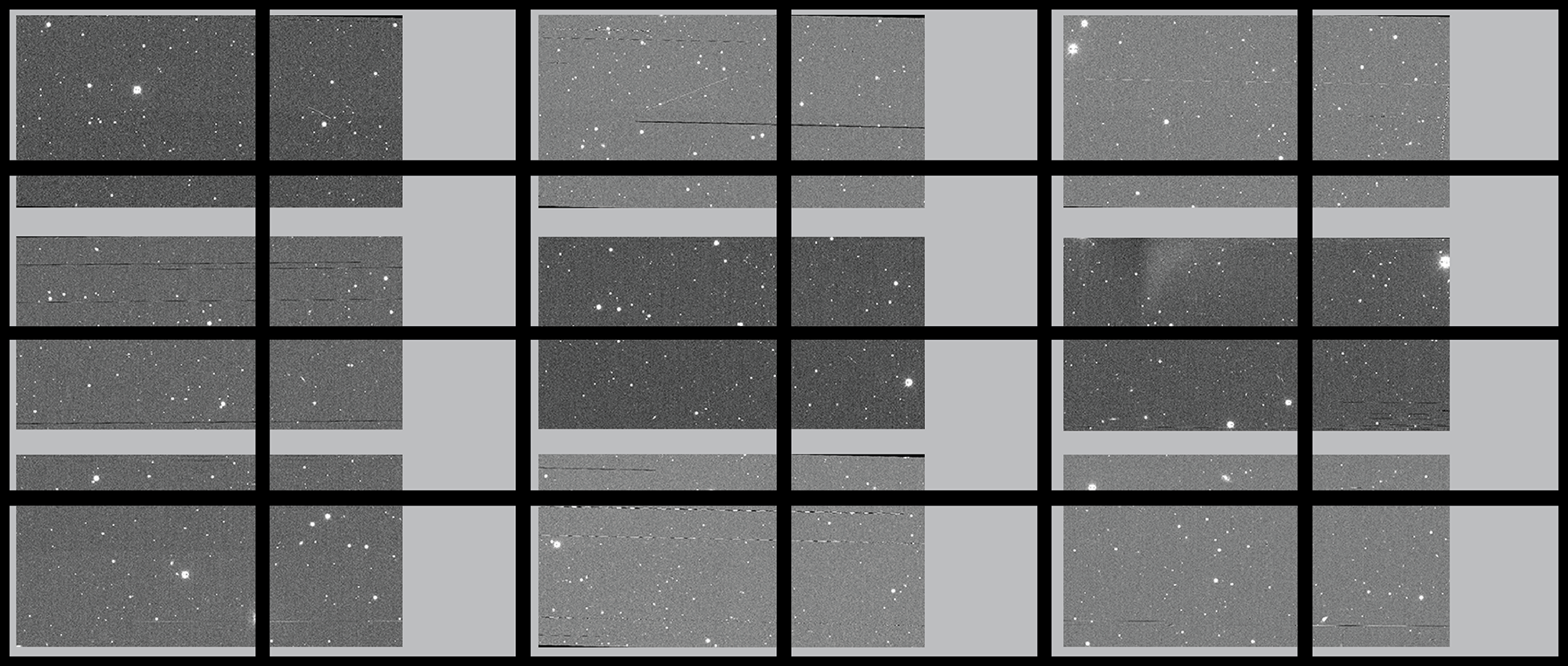}
\caption{During O1, in order to avoid any image size reduction, the best image configuration for the TDW was 3 $\times$ 3.  This provided clear separation between images but also meant that each image spread across 4 screens. The bezels did not obscure any image pixels.}
\label{fig:tdw9images}
\end{center}
\end{figure*}

\section{O1: the December 2015 campaign}
\label{sct:run3}

To address the shortcomings of the pilot program (Section \ref{sct:phase1}), a new workflow was developed for O1.  This included the use of an automated candidate identification pipeline and an improved visualisation process to manually review the CCD images.

\subsection{Automating candidate selection}
The simultaneous multi-wavelength imaging strategy and real-time optical imaging analysis component of {\em Deeper, Wider, Faster} is as follows:
\begin{enumerate}
\item {\bf Data collection and transfer:} The DECam electronics provides a 20s readout time for the entire set of 62 CCDs.  The {\em Deeper, Wider, Faster} program chose to take a continuous stream of 20s exposures to provide the fastest cadence to search for fast transients, while maximising survey depth and time on sky. Fields are simultaneously observed for 1-2 hours by several observatories, with the time on field constrained by the coincident visibility by DECam in Chile and Parkes and Molonglo in Australia. As a result,  around 100-200 DECam optical images are aquired per field. Image files are 1.2 GB (uncompressed), but total data increases by 3-4 times during processing.  Images and processed files are stored on Swinburne University's Green II Supercomputer facility.  While there are 62 science CCDs in the DECam array, only 59 were usable during O1 as two CCDs were non-functional and one had a damaged amplifier and could not be calibrated. 
These 59 CCD images are referred to as a {\em batch} for the remainder of the paper.


JPEG2000 compression was performed at CTIO in order to compress the data sufficiently to expedite the transfer (Vohl et al. {\em in prep}). For this purpose, we modified the {\tt KERLUMPH} software \citep{vohl2015} to convert files from the FITS format\footnote{http://fits.gsfc.nasa.gov} into the JPEG2000 (ISO/IEC 15444) format \citep{jpeg2000}.  The level of file compression was determined on-the-fly to keep transfer time reasonable while maintaining sufficient information to achieve the science goals.  Because several of the subsequent processing steps did not support JPEG2000, the images were converted back to FITS. Data transfer from CTIO to Swinburne University took between $\sim 3$ and $\sim 15$ minutes per batch of images during O1, and $\sim 1$ to $\sim 5$ minutes for O2. On reaching Swinburne, the images are uncompressed and processed (Andreoni et al. {\em submitted}).

\item {\bf Initial processing:} Data were processed in stages using eight reserved nodes on the Green II supercomputer.  
\begin{enumerate}
    \item Individual CCD images are calibrated using parts of the PhotPipe  pipeline \citep{rest2005}. 
    \item The {\em Mary} pipeline (Andreoni et al. {\em submitted}) is used to coadd, align, and subtract the images, and to automatically search for transients. {\em Mary} identifies CCD artefacts and poor subtractions to reduce a sample of several thousand initial detections (across all CCDs) to a few tens of objects. 
    \item Finally, {\em Mary} generates products for visualise inspection such as postage stamp images (varying between 80px and 120px per side) and region files identifying the nature of known variable and other sources from catalogs to assess potential candidates to follow up. 
\end{enumerate}

\item {\bf Visual inspection:} At the same time as the {\em Mary} pipeline was extracting potential candidates, the full-frame difference images were also viewed in their entirety. While the preference was to maintain the FITS format, the requirements of the TDW necessitated converting the images to JPEG in order to display in the Scalable Amplified Group Environment (SAGE2)\footnote{\url{http://sage2.sagecommons.org/}} environment. For O1, it was thought that this was more important to achieve than the use of FITS compatible software (see Section \ref{sct:tdwsoftware} for more detail), however for O2, with the more developed pipeline for eliminating unwanted candidates, the benefits of FITS was more important.  This corresponded with requests from the O1 inspectors to reconfigure the TDW with only two rows for ease of use during O2.  The use of FITS files enabled demarcating software identified candidates on the full CCDs, as well as known variable objects, known CCD crosstalk, and other information.  This approach provided a visual confirmation of the efficacy of the {\em Mary} pipeline, as well as the opportunity to find targets possibly missed by the pipeline.  Other problems with potential candidates that could fool the automatic system -- but hopefully not a trained astronomer -- include amplifier crosstalk and CCD defects  (see Sections \ref{sct:crosstalk}).
\end{enumerate}


\subsection{Room Configuration}
\label{sct:roomconfig}
\begin{figure*}
\begin{center} 
\includegraphics[width=16cm, angle=0]{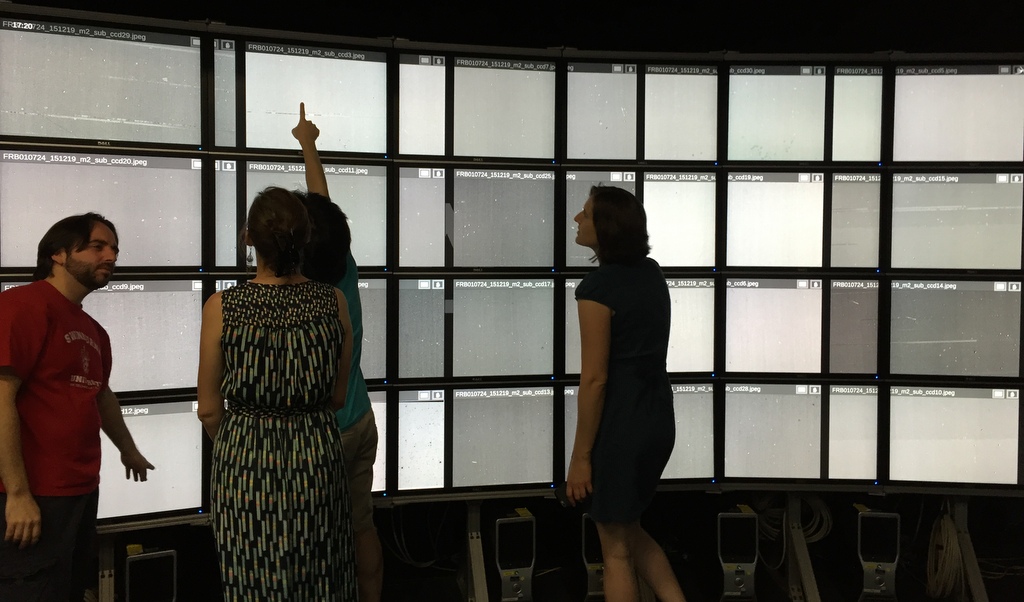}

\includegraphics[width=16cm, angle=0]{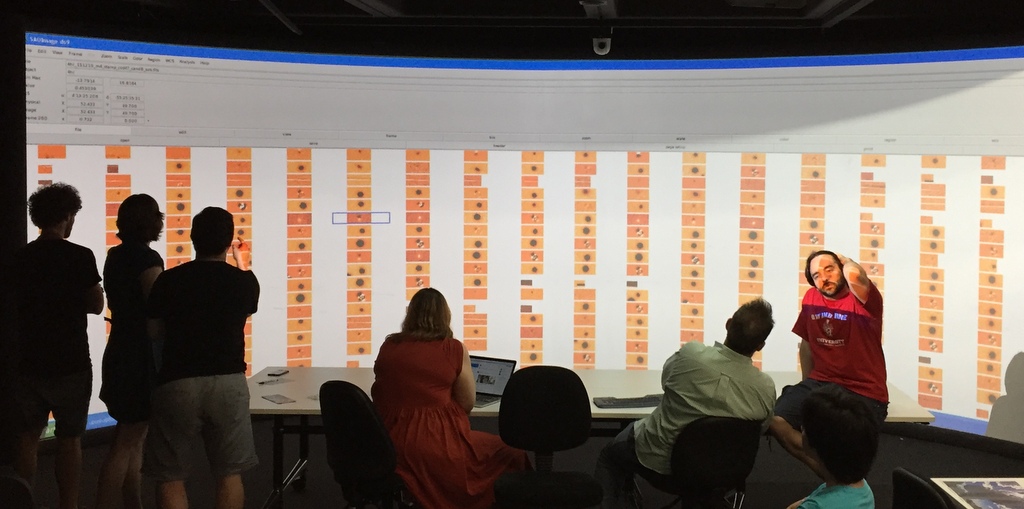}
\caption{(Top) The OzIPortal TDW with images displayed in 4 $\times$ 5 configuration during O1.  Several configurations were tested but the 3 $\times$ 3 configuration was deemed most suitable. (Bottom) A large number of candidates, with science images and subtractions, shown as postage stamps, can be inspected at once by several researchers, and shared with anyone in the room.  This was particularly useful in supporting novice inspectors. }
\label{fig:tdw}
\end{center}
\end{figure*}

The Advanced Immersive Environment at the University of Melbourne was chosen as the base of operations for O1.   It offered access to a 98 Megapixel TDW, a large-area curved projection screen, and table-top work spaces for the  team members to bring and use their own devices.  Moreover, with around 100 square metres of floor-space, there was ample room for the team to move, work and collaborate effectively.

During O1, the room was configured with two principal enhanced display technologies -- see Figure \ref{fig:layout} and Table \ref{lbl:hardware}.  One end of the room was occupied by the TDW and was used to display the processed difference images in JPEG format.  Each CCD image was 4000$\times$2000 pixels and the TDW display area was 15360$\times$6400 pixels.  In order to optimise the 6$\times$4 display configuration and ensure the images were shown at native resolution, the images were presented in a 3$\times$3 matrix to allow clear space between the images, and reduce the need to use the uppermost region of the TDW.  It was necessary to have the images appear across 4 screens with no pixels hiding behind the bezels, as shown in Figure \ref{fig:tdw9images}.

At the opposite end of the room was the curved projection screen, which was used to show the postage stamps images of the software-detected candidates. Typically around one hundred of these candidates were displayed simultaneously across the 6.9m $\times$ 2.2m display using SAOImage DS9.  Other applications could also be displayed alongside the DS9 window, such as IRAF, to assist in the evaluation of the candidates.

Other operations were positioned between these two displays to allow easy observation from the process facilitators.

\subsection{Tiled Display Wall Software}
\label{sct:tdwsoftware}

The SAGE2 software was used to manage the display windows being presented on the TDW.  In this client server model, the SAGE2 Head Node acts as a HTML5 web server, with the ``clientID'' tag being used to specify which window is to be streamed to the client.  For example, with the head node running the ``node.js'' based service, a Firefox window is launched on a tile display node, and directed to a particular URL for that frame.  This environment was chosen as it made it possible to script the loading and display of the CCD images on the TDW, as well as log the time taken to do so.  It also enabled easy review of the individual images at much larger scale for close scrutiny, as the images could be expanded across the entire display if required.

\subsection{Workflow}
\label{sct:workflow}

As a batch of images arrived at the Green II Supercomputer, the {\em Mary} pipeline produced a collection of potential candidates for display on the curved projection screen.  In parallel to this process, {\em Mary} generates template-subtracted FITS files which are converted to JPEG.  This conversion was necessary for O1 but abandoned for O2; see Section \ref{sct:probstdw} for more detail.  These were transferred directly to the head node of the TDW and visually searched for potential candidates (see Section \ref{sct:candidate}).  

Python scripts were used to present the images on the TDW running the SAGE2 interface.  The display script automatically loaded and positioned nine images on the display initially and as the researchers completed the inspection of an image, it was replaced with a new image.  After the initial images were loaded, an operator monitored the progress of the inspectors, and manually advanced the script to load the next image when it was clear the inspector had moved on to a new image.  This process continued until the full set had been inspected.  See Figure \ref{fig:tdw} (Top).

The images were presented in columns with an image placed at the bottom of each column and progressively moving up in rows.  As each column was assigned to a researcher, the presentation order was intended to reduce the wait time for each researcher to start their inspection task.  In this way, each researcher was presented an image in their assigned column before anyone else received a second image.  The choice to start with the bottom row was decided by the people inspecting the images as preferable to loading top down.

As images were inspected, candidates of interest were flagged for follow up, with approximate locations noted.  Initially, interesting candidates were recorded using paper (as this was a natural reaction) but was then moved to the whiteboard as seen in Figure \ref{fig:wholeroom}. These targets were then inspected on a standard laptop computer running SAOImage DS9, and on the curved screen, also running DS9.  Where necessary, images could be recalled to the TDW for verification and comparison.

Several bash and python scripts were developed to expedite the workflow.  These include such things as moving a completed batch of jpegs to a storage folder to make way for the next batch, or starting or stopping the TDW nodes.

\subsection{Training the image inspectors}
\label{sct:participants}

Several of the team members had never worked with a TDW or curved projection screen before, while others had extensive experience.  Roles ranged from observing the use of the display technologies while working on their own tasks, to those who worked exclusively with the TDW and/or curved screen.  See Figures \ref{fig:wholeroom} \& \ref{fig:tdw} for examples of the displays in use during O1.

Introducing the candidate identification/rejection process required a short training session for the image inspectors. \citet{meade2014} found that using a TDW was an unfamiliar experience for most people and without an introduction, it was unlikely to be particularly useful. However, a short explanation of physical navigation, i.e. physically moving your body to achieve the equivalent of panning and zooming, improved the experience and efficacy of using a TDW.

This orientation process was augmented for O1 by using sample images showing examples of potential candidates, as well as examples of system or processing errors, such as badly subtracted images and crosstalk.  The collaborative environment meant volunteers could be trained ``on the fly'', which was very useful considering the dependence on volunteers with varying availability.

\begin{figure*}
\begin{center} 
\includegraphics[width=16cm, angle=0]{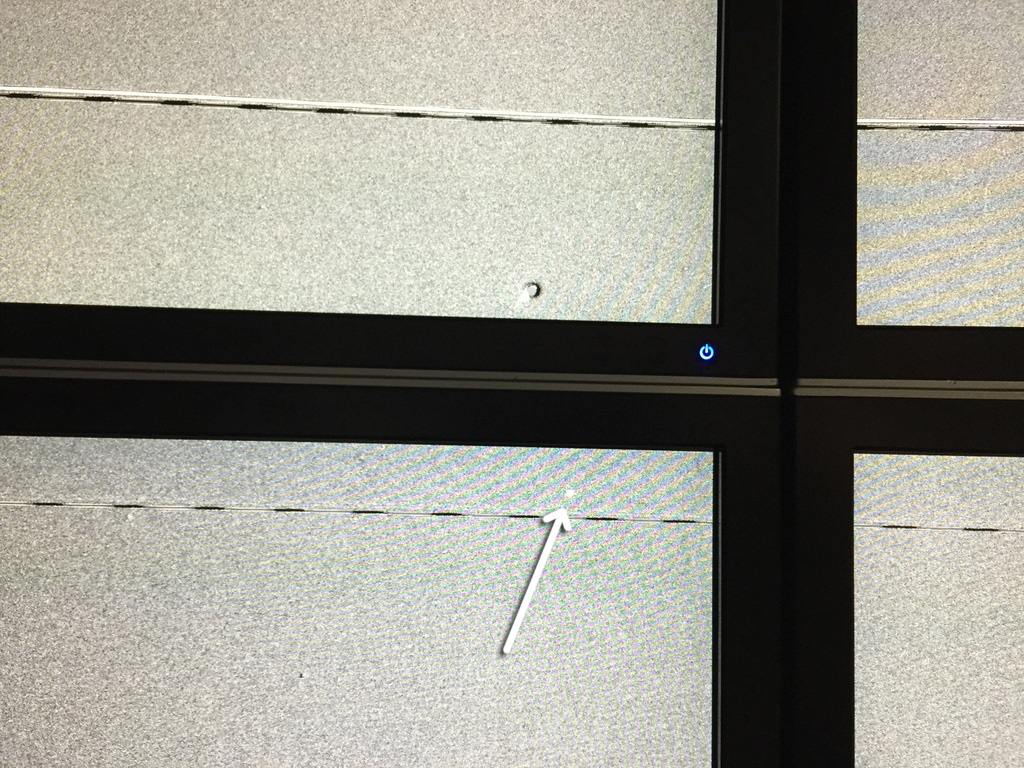}
\caption{An example of a potential candidate on the TDW that meets all the necessary criteria for closer inspection and possible follow-up with other telescopes.}
\label{fig:candidate}
\end{center}
\end{figure*}

\subsubsection{Potential Candidate}
\label{sct:candidate}
Potential candidates are expected to appear as 2D Gaussian point sources in the images and (roughly speaking) appear as small, round objects with soft edges and no black (negative flux) artefacts that could indicate poor subtractions of non-transients or CCD effects such as bad pixels or column subtractions.  If a potential candidate met each of these conditions, they were usually corroborated by other researchers and then flagged for more detailed inspection, with approximate coordinates noted -- see Figure \ref{fig:candidate}.

\subsubsection{Amplifier crosstalk and CCD defects}
\label{sct:crosstalk}
\begin{figure*}
\begin{center} 
\includegraphics[width=16cm, angle=0]{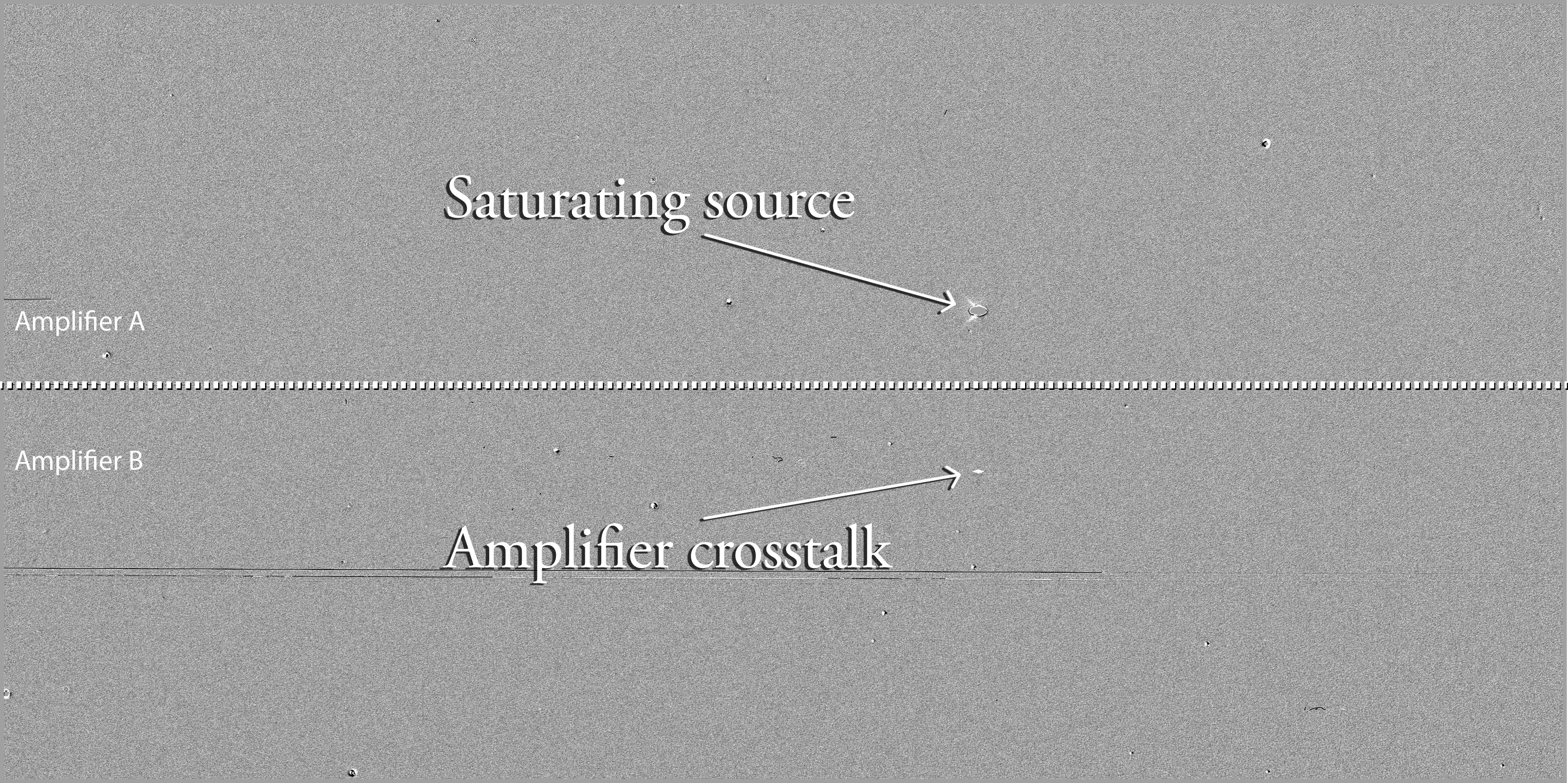}
\caption{Each CCD has two amplifiers reading out each half of the image.  Sometimes this will result in a crosstalk image of a saturated source from one amplifier to the other. }
\label{fig:crosstalk}
\end{center}
\end{figure*}

Each CCD has two amplifiers that can create artefacts when processed by the operating system electronics.  When a source in the region of amplifier A saturates, it creates a crosstalk image in the region of amplifier B, equidistant from the line joining the amplifiers.  Potential candidates that had a clear counterpart on the opposite site of the image could be eliminated from consideration, such as shown in Figure \ref{fig:crosstalk}.  

\begin{figure*}
\begin{center} 
\includegraphics[width=16cm, angle=0]{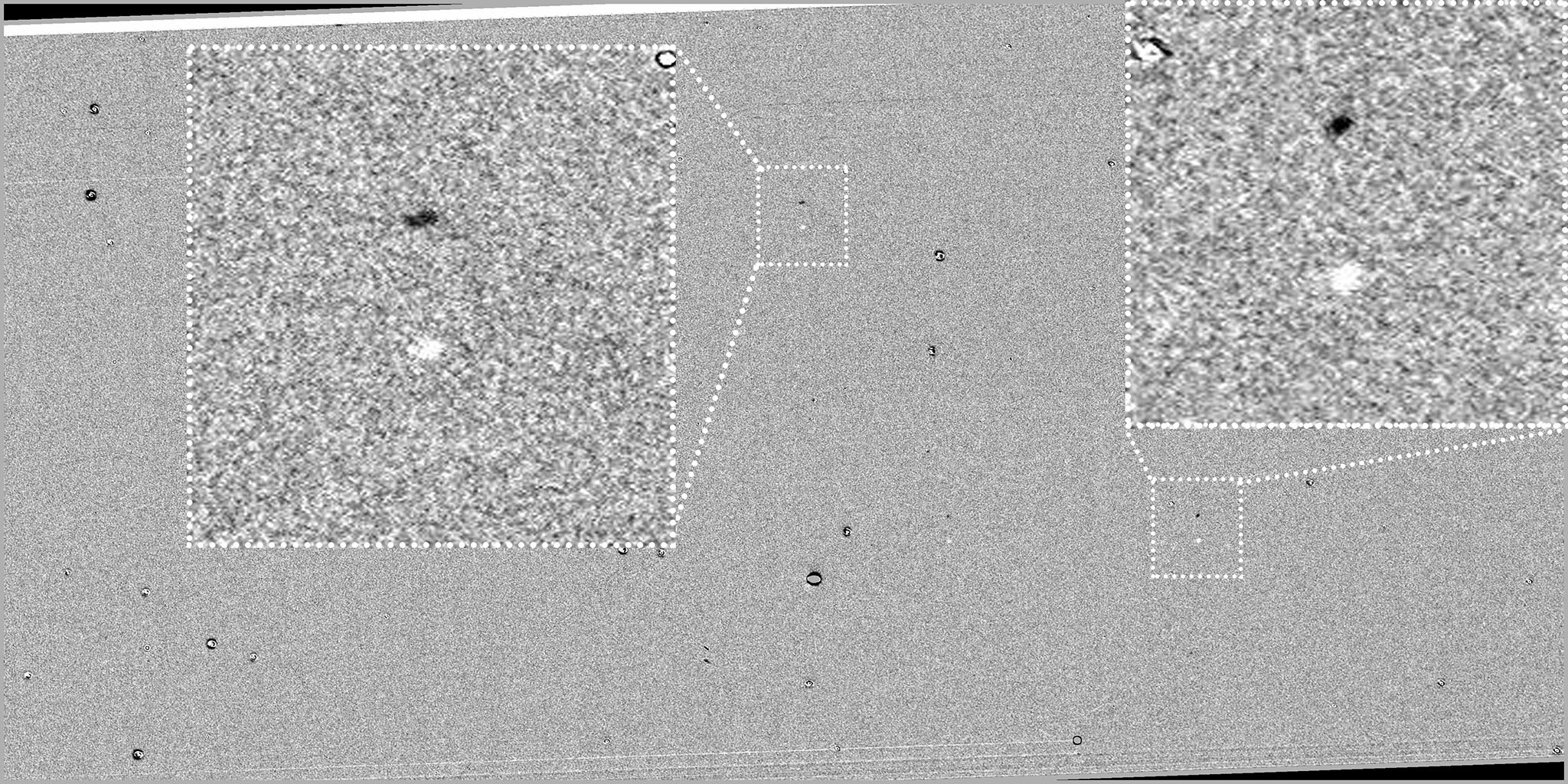}
\caption{When several potential candidates show a negative partner  offset by a regular amount, the potential candidate can be eliminated from consideration.}
\label{fig:offset}
\end{center}
\end{figure*}

Occasionally what appears to be a potential candidate shows a negative partner observed at the same  offset as other potential candidates within the image, as shown in Figure \ref{fig:offset}.  While the precise nature of this effect is unknown, it is likely an artefact of the DECam CCDs and the fast data processing, and not celestial phenomenon. Fortunately the display ecology helps easily identify the effect that could be missed by other conventional identification techniques.

\subsubsection{Time spent on tasks}
\label{sct:quantitative}

The workflow described above was used by the {\em Deeper, Wider, Faster} team over the six nights of O1.  The team assembled from around 12pm and prepared for on sky observations at 3:00pm until 7:30pm.  The direct measure of image loading time on the TDW was able to be tracked by a log generated by the script for displaying the targets.

The transfer of converted JPEG images from the Green II cluster at Swinburne University in Hawthorn to the SAGE2 head node located at the University of Melbourne's Queensberry Street Data Centre in Parkville, did not contribute significantly to the workflow and the transfer time was not tracked.  
















An image display control script was used to populate the TDW with images as quickly as possible, positioning the first 9 images in a 3$\times$3 matrix. As soon as an image was available, inspection started. 

The aim of the control script was to ensure the participants always had a new image available when they were ready to move on.  The initial loading time for the images was quite consistent, with an average time for the first 9 images of 54.9 seconds.  Each column had an image within 20 seconds, which includes additional scripted delays such as clearing the TDW (2 seconds), loading and positioning images in each column (2 seconds to load and 2 seconds to place for each column, totaling 12 seconds). These delays were built in to the script to avoid race conditions, that is, where a compute process attempts to complete two or more tasks at the same time and fails, at the head node.

While the time to completely review a full batch of images was not formally recorded, Table \ref{lbl:loadtime} shows the duration of the TDW image review process taken from the first image loaded to the last image loaded for that day, and the number of images reviewed, as logged by the control script.

\section{Evaluation of the collaborative workspace for O1}
\label{sct:results}

\subsection{Expectations}
\label{sct:qualitative}
Before O1 began, several members of the {\em Deeper, Wider, Faster} team reflected on the role that an alternative display ecology might have on overcoming the limitations of the pilot program.  The comments here refer only to the TDW, as the use of the curved screen for O1 had not been confirmed at the time.  Four broad themes emerged.


\textbf{Throughput:} Utilising multiple astronomers to inspect the images in parallel should improve the throughput of the images in a set.  By having the images automatically loaded and positioned on the TDW for the astronomers, there should be no wait time once the first image is available for inspection.  This assumes that it is quicker to load new images than to inspect an image.  Each astronomer can complete their images and should time permit, they can easily assist others.

\textbf{Rapid corroboration:} With astronomers inspecting images side-by-side, there is the potential for rapid corroboration of a suspected candidate.  An astronomer can easily leave an image being inspected to assist a colleague nearby to determine the viability of a candidate.  When complete, the astronomer can easily return to their own image.  Because the researchers are in close physical proximity, this can happen very quickly.  The short delay in inspecting a particular image should not make it difficult to return to the image and pick up where the astronomer left off.

\textbf{Native resolution:} A thorough inspection of each image is necessary as the potential candidates are likely to be represented by only a few pixels.  Having the images shown at full resolution should reduce the possibility of overlooking potential candidates that might be missed due to subsampling caused by scaling, or when panning and zooming.  This should also help rapidly identify artefacts and thereby reduce time spent on non-candidates.

\textbf{Workflow optimisation:} The use of an advanced display such as the TDW should improve the overall workflow and help design future workflows that are optimised for speed and accuracy.  It should also help refine the pipeline in the identification of candidates for the future.

\begin{table}[]
\centering
\caption{The image display control script was used to log the start and end times of image loading during O1.  The shorter duration on the 23rd of December was due to problems with DECam that limited observing time.}
\label{lbl:loadtime}
\begin{tabular}{lll}
\textbf{Date} & \textbf{Duration} & \textbf{Images}    \\ \hline
2016-12-18          & testing & N/A   \\
2016-12-19          & 2h 30m & 424   \\
2016-12-20          & 2h 29m & 303   \\
2016-12-21          & 4h 45m & 494   \\
2016-12-22          & 4h 22m & 267   \\
2016-12-23          & 2h 39m & 341   \\
\hline
\end{tabular}
\end{table}



\subsection{Impact of the Tiled Display Wall}
At the conclusion of O1, the {\em Deeper, Wider, Faster} team again reflected on their experiences, this time with both the TDW and the curved screen.  While successfully meeting the expectations (Throughput, Rapid corroboration, Native resolution, Workflow optimisation), additional themes were identified.

\label{sct:tdwimpact}
\textbf{Candidate rejection:} Crosstalk artefacts are due to the dual amplifiers for each CCD.  When this occurs a potential candidate can be eliminated from further consideration because it is being generated by a non-candidate in the other amplifier.  The observing strategy we adopted in O1 avoided performing dither patterns in order to maximise the continuity of sampling each part of the CCDs.  As such, we uncovered the extent of this effect but, at the time, it was difficult to consistently anticipate crosstalk locations.  Non-candidates that would have been otherwise discarded might appear as potential candidates in the reflected part of the image.  The postage stamps themselves are not large enough to show evidence of this effect, however, it is quite easy to identify this phenomenon when looking at the whole image on the TDW.

\textbf{Quality control:} Other errors such as CCD defects, CCD processing problems from the real-time pipeline or telescope tracking or guidance problems are far more obvious on the full resolution images displayed on the TDW.  When time is of the essence, rapid identification of faults is essential to avoid wasteful delays and prevent rapid-response telescopes triggers on non-celestial sources.

\textbf{Missed discovery:} As with any automated system, it needs specific criteria in order to make a selection.  While this does not mean an entirely new phenomenon cannot be discovered, it does open the possibility of missing something that might catch the eye of a trained astronomer. 

The sheer volume of data being collected from astronomical instruments these days mean it is essential to exploit automatic processes wherever possible, as typically there is simply too much information for human eyes to sift through in a meaningful time.  The best option is the combination of automatic processes and manual inspection.  As the automatic processes become more mature, they reduce the pressure on the manual processes, though it is hard to imagine if full discovery space can ever be fully automated.  In the context of unbiased searches for fast or exotic transient events, the combination worked extremely well, with both the curved screen and TDW inspection processes being used to great effect to support each other.  

\textbf{Throughput:} Images displayed on the TDW are able to be inspected far more quickly than was possible in the previous run of the experiment.  Parallel inspection with several astronomers working on separate images significantly speeds up the process, with one observer estimating around 50\% improvement in efficiency of detection confirmation or rejection of candidates.

\textbf{Native resolution:} The objects of interest are small, usually representing less than 0.08\% of the image area.  They are circular and have a soft edge i.e., a point source, thereby having a 2-D Gaussian-like shape, yet this is often lost when the image is subsampled, such as when viewed full screen on a display of lower resolution than the image.  On such a screen, many more objects appear to have this profile until they are zoomed into, when they can be seen to be not circular, or have hard edges or other artefacts not apparent before.  The TDW (and indeed any display capable of displaying images at native resolution) eliminates the need to zoom in, and so speeds up the rejection of non-candidates greatly.

The TDW encouraged whole body movement to scan images rather than just with eyes.  This maintains the scale of an object in the context of the image which is difficult to match when panning and zooming on a standard desktop display.

\textbf{Human factors:} Another benefit of the TDW was the ability to recall images for the purpose of comparing epochs.  To perform effective transient candidate detection, it is necessary to recall images from other epochs for comparison.  During O1, the automatic process was not designed to efficiently crosscheck every candidate in previous images that the manual process was able to perform.  This feature was added for O2, and complemented by the online logging tool (see Figure \ref{fig:spreadsheet}.  Not only did this identify several interesting events worth following up, it is also invaluable for maturing the automatic process for future observations.  Giving objects `running IDs' has since been employed to allow precisely this sort of temporal tracking for subsequent runs.   

\subsection{Problems with the Tiled Display Wall}
\label{sct:probstdw}
While aiding with the throughput, rapid corroboration, and overall experience, the TDW posed some logistical and operational challenges.

The physical height of the TDW made it difficult to see the upper regions of the images in the top row for some inspectors.  The lowest regions of the bottom row also presented some difficulty as they required the inspector to bend or squat, which became uncomfortable after several hours of moving up and down through the images.  This could be improved by reconfiguring the TDW into a 12$\times$2 configuration, which would redeploy the top and bottom rows, providing a more comfortable fit with the average viewing height.  This new configuration would allow additional columns of images, making it easier to include additional inspectors.  The practicalities of changing the configuration made it too difficult to employ during O1 but was implemented for the July/August 2016 UT (O2) campaign.
    
As the TDW is necessarily made up of many smaller screens, it is impossible to avoid screen edges.  While it is possible to purchase screens with negligible bezel (screen edges) size, these are very expensive.  The screens used in this TDW have bezels of 20 mm, making a combined bezel width of 40 mm, and sometimes more due to slight gaps between the screens themselves. As image resolution exceeded the screen size, each image spanned four displays (see Figures \ref{fig:tdw9images} and \ref{fig:candidate}), with a break in the image at the screen edge. \citet{meade2014} showed the practical and psychological impact of the screen bezels on an observer is typically small, however inspectors reported it can be distracting.  When potential targets that lay within a few pixels of the break in the image were encountered, the image could be shifted slightly to place the candidate in question in an unbroken region of the display.  This however requires additional time, but fortunately happened only a few times, none of which resulted in a positive candidate selection.  Reducing the physical size of the bezels would reduce this problem.

The depth of the bezels to the screen surface also meant that for the top row of screens, the outward protrusion of the bezel itself was obscuring pixels at the bottom edge of the screen, as seen by someone looking upward.  Reducing the depth of the bezel and/or reconfiguring the tiles to reduce the need to look upward as much would reduce the possibility of missing candidates.
    
It was necessary to convert from FITS to JPEG in order to display on the TDW due to format restrictions of the SAGE2 software.  This added an additional step to the workflow which, while relatively minor, became a tripping point on several occasions.  Minor human-generated mistakes, such as beginning a transfer before the full set had been converted meant the process had to be repeated to pick up missed images.  Transferring was initiated manually and on multiple occasions saw a set of images transferred too soon, overwriting a set of images during inspection.  If the TDW could handle FITS images directly, possibly with alternative software, then the transfer step could more easily be automated and would reduce the potential for human error.  Such an approach has been successfully tested by \citet{pietriga2016} with the FITS-OW software, but that application is still in development.  An alternative would be to configure the TDW in subgroups of 2 $\times$ 2 screens connected to a single computer, thereby allowing the direct use of SAOImage DS9 to display the FITS images.
    
The process of loading, resizing and moving images was slower than expected due to race conditions at the SAGE2 web server.  In order to avoid this, short delays of 1-2 seconds were built into the scripts to ensure a response from the web service.  Once loaded, manual movement and scaling of the images was acceptable.  These delays were added in situ to cope with problems as they occurred.  While error trapping would have negated the need to incur delays on each load and move command, the time to develop such a solution was not deemed useful during O1, as the cumulative delays were only in the tens of seconds over a batch of images.
    
SAGE2 did not provide a convenient way to flag potential candidates and note their coordinates within an image.  The number of promising candidates were relatively few and were relayed to the analysts by identifying the CCD and approximate location of the candidate either verbally or via notebooks and use of the whiteboard.  While not ideal, this did not cause a major problem as it was typically done to verify a potential candidate within a current run, so the coordinates were reasonably well known.  However, an online logging tool shared in real-time has been developed for subsequent campaigns.  This allows inspectors to use laptops and mobile devices to log potential candidates in situ, resulting in improved reporting consistency and tracking of review outcomes.
    
    
While the process of inspecting the images was entertaining and engaging, after closely inspecting several hundred images, the observers did become tired.  This was due to the mental demands of being thorough and the physical requirement of standing in front of the TDW for several hours.

\subsection{Impact of the curved projection screen}
\label{sct:cpdimpact}

When viewing the postage stamps of potential candidates produced by the {\em Mary} pipeline, the curved projection screen provided a more suitable display surface than the TDW.  The curved display has a resolution of 3440 $\times$ 1200 (due to the 400 pixel blending region) over a physical display surface of 6.9m $\times$ 2.2m. 

The immersive nature of this display enhanced the experience for the researchers as several reported feeling more engaged with the information being presented.  Driven by a single computer, the curved screen produces a more ``desktop" like experience that could be easily viewed by everyone at once, especially those across the room examining the full CCDs.  There was a faster level of responsiveness when compared to operations performed through SAGE2.  Without the physical presence of bezels on the TDW, the image blending of the two projectors provided an uninterrupted display area, and so all images remain unbroken.  
    
Not only could many postage stamps of potential candidates be viewed simultaneously ($100$ images was typical), it is also possible to have other applications running alongside the SAOImage DS9 software.  The X11 applications were forwarded from the Green II cluster.  The bandwidth provided between sites (Swinburne University to the University of Melbourne) was adequate to operate the application with negligible lag.  Other software such as IRAF and multiple terminal windows were also forwarded from Green II, and displayed alongside the candidates being presented by DS9. Having all the necessary information readily available and easily viewable by several people demonstrates the utility of the environment.

At $15.18$m$^2$ the actual surface area of the curved screen is much larger than the $6.17$m$^2$ of the TDW. This increased physical size made it easy for several people to collaborate on the same content at once.   Moreover, the screen's curvature meant that content at the far edges was less horizontally compressed (from the central viewing point) than with a flat display of equivalent size.

\subsection{Problems with the curved projection screen}
\label{sct:probscpd}
As with the TDW, the curved projection screen posed several challenges with regards to its use, suitability, and display qualities.

When compared with the TDW, the curved projection screen has much lower resolution and contrast.  The increase in area afforded by the curved screen was slightly counteracted by the lower pixel density: $15.9$ pixels/$mm^2$ to $0.27$ pixels/$mm^2$ respectively. Despite the lower resolution, the curved projection screen was ideal to display the postage stamps whose resolution is comparatively very low, thus, the pixels were resolved.  Similarly, the text windows and graph displays, while not perfectly sharp, were quite adequate for the task. 

While approaching the screen surface did result in shadows cast by the front projection system, this did not discourage the astronomers from getting very close to the screen to discuss objects of interest.   There was a slight impact from the in-room lighting.   For safety and general usability of the collaborative workspace, some lights were required to be on during the observation to facilitate people moving around the room. While the spill from the overhead lights was minimal, reducing it even further would have been desirable. 
    
The lack of suitable drivers for the hardware used by the curved projection screen dictated the use of Microsoft Windows XP SP4.  Rather than locating and installing Windows versions of the preferred Linux applications required, X11 forwarding was tested (with Putty\footnote{\url{http://www.chiark.greenend.org.uk/~sgtatham/putty/}} and XMing\footnote{\url{https://sourceforge.net/projects/xming/}}) and found to perform very well.  The applications were being forwarded from Green II at Swinburne, where the image data was stored and the cluster processing occurred.  
    
As a single large display space, the curved screen functioned as a standard, albeit very large, desktop computer.  A useful capability would be the ability to drive the display from another computer with a pre-configured environment more suited to the task, with drive paths and device drivers already installed.   Also being able to have multiple people working independently on the same screen but in separate windows, with their own keyboard and mouse control, would greatly increase the versatility of the environment.  We suggest to introduce an intermediate step to achieve this would be to use screen sharing, with researcher laptops being replicated on the curved screen.

\section{O2: the July/August 2016 campaign}
\label{sct:run4}

\begin{figure*}
\begin{center} 
\includegraphics[width=15cm, angle=0]{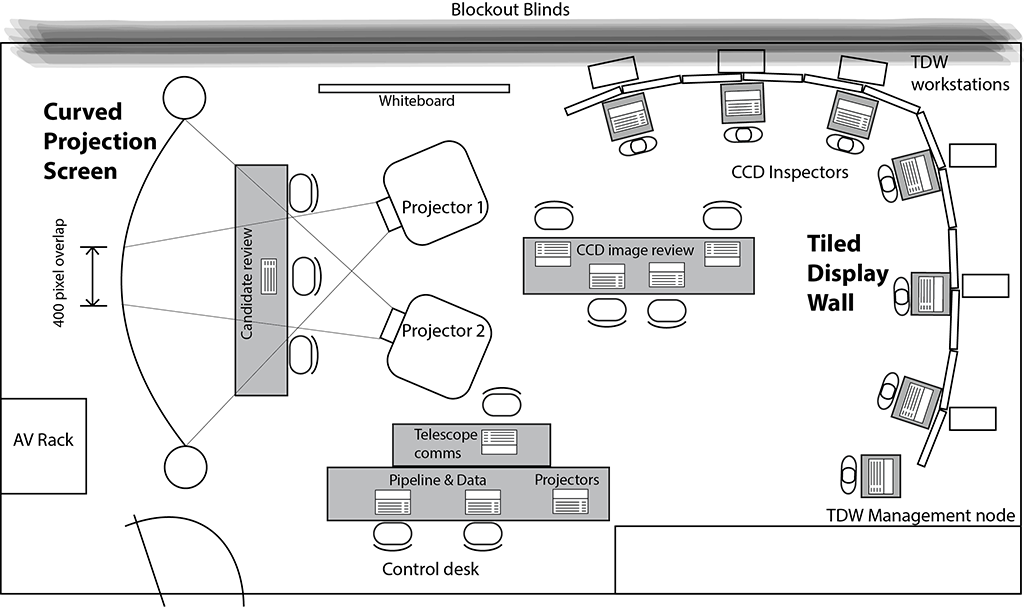}
\caption{Updated layout of the Advanced Immersive Environment at the University of Melbourne.  The curved screen for reviewing the {\em Mary} candidates remained unchanged from O1 to O2.  The TDW was broken into 6 workstations with 2 $\times$ 2 tiled screens, and space for a users laptop.  The central desk was also rotated to facilitate better movement between work areas. }
\label{fig:layout2}
\end{center}
\end{figure*}

\begin{figure*}
\begin{center} 
\includegraphics[width=15cm, angle=0]{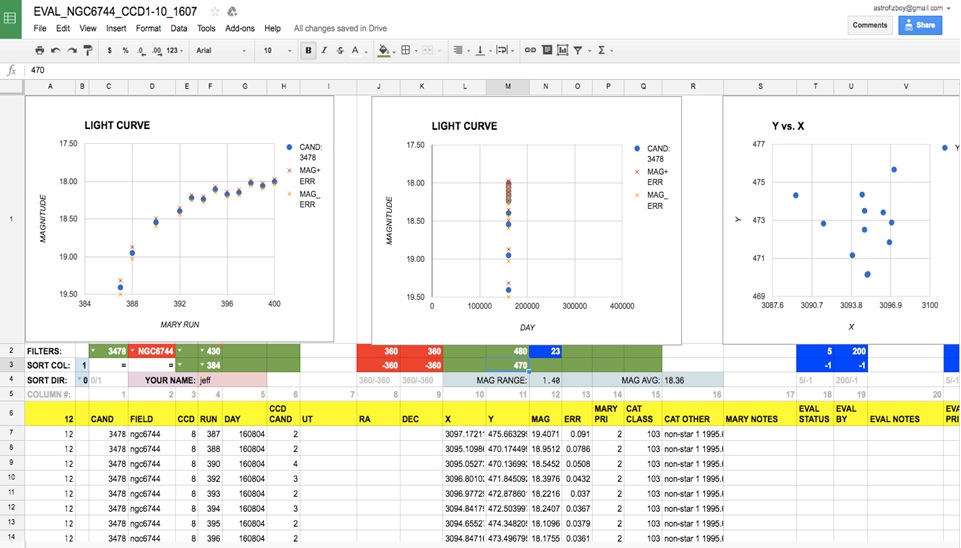}
\caption{The {\em Deeper, Wider, Faster} online logging tool allowed the inspectors to track the light curves of the potential candidates, their postage stamp images, and candidate positions, magnitudes and other information.  From this tool, the inspectors could report targets of high priority to the principal reviewer for trigger consideration. However, the tool did not have the capability to show the full CCD images.  This capability has been added in a later version of the tool. }
\label{fig:spreadsheet}
\end{center}
\end{figure*}

Operating from the 26 July 2016 until the 7 August 2016, with between 5.5 and 11 hours each observing session, O2 adopted several improvements in workflow. 

\begin{itemize}

\item{\bf Display ecology}: The TDW was reconfigured, eliminating the top and bottom rows and spreading the screens out into 6 workstations with 2 $\times$ 2 screens, with each workstation computer operating largely independently, though sharing a file system with the other workstations.  With 5120 $\times$ 3200 pixels, these workstations were able to display the 4096 $\times$ 2048 pixel images at native resolution, and provide sufficient screen real estate for the SAOImage DS9 toolbar.  The addition of desks in front of the screens provided inspectors with a place to use their laptops to access the online spreadsheet, however, they could still stand if required.   See Figures \ref{fig:layout2} and \ref{fig:wholeroom2} for the updated layout.

The new configuration eliminated the physical observation problems associated with the TDW for O1.  With the top and bottom rows removed, the screen height was more consistently comfortable (though not customisable to individuals).  This also effectively removed the bezel depth occlusion problem described in Section \ref{sct:probstdw}, as the inspectors were easily able to reposition themselves to eliminate the issue.

\item {\bf Software:} Using DS9 allowed the inspectors to load FITS images directly from the TDW head node. Along with the FITS images, automatically generated DS9 region files were available for overlay on the images.  These regions included the persistent Candidate ID numbers, making the process of identifying them within the full resolution image much simpler for the inspectors.

\item{\bf Event logging:} Using an online spreadsheet to log and track potential candidates, including their real-time light curves, the requirement for immediate inspection of the full CCD subtraction image was mitigated.  Instead, individual or groups of images could be called up for review if they had already been identified by the {\em Mary} pipeline and logged in the spreadsheet, as seen in Figure \ref{fig:spreadsheet}.
\end{itemize}

The rest of the workflow remained relatively unchanged from O1. This preserved the collaborative and training benefits of the workspace from O1. While other optimisations to the detection pipeline were made, these did not affect the overall workflow significantly.

Improvements to the workflow from both the {\em Mary} pipeline and the display ecology resulted in significant outputs for O2.  Several triggers were sent during the run to Gemini-South and SALT, with more than 50 targets identified for subsequent follow-up with Skymapper and the Zadko Telescope. Hundreds of candidates received spectra and follow-up imaging and tens of thousands of candidate variable and transient objects were detected.

As the principle objective of the display ecology had been established in the planning and execution during O1, no formal attempt was made to conduct an additional review for O2.  Instead, the achievable recommendations from the inspectors after O1 were implemented and subsequent workflow improvements developed organically in response to the new configuration.  The positive response from the inspection team was unanimous in supporting the need for the display ecology.

\section{Discussion}
\label{sct:discussion}

O1 operated on sky for approximately 4.5hrs per day, with visual inspection continuing for up to an hour longer, over six consecutive days where the workflow was continuously refined.  O2 operated for between 5.5 and 11 hours, with additional time for visual inspection, over 13 consecutive days.  As a mark of a successful endeavour, the focus shifted away from the workflow to the survey itself. Future refinements based on the experience acquired during each of these runs will will greatly improve the chances of successful real-time, fast follow up with additional telescopes.


The workflow adopted for O1 and O2 alleviated many of the frustrations associated with the pilot campaign.  Establishing a functional display ecology with the ability to display all the relevant content and context simultaneously improved the confidence of the observers that they were getting all the necessary information to make the appropriate decision about candidates.  The collaborative environment also improved the observers' experiences during the survey, to the point that reverting to the previous workflow could compromise the purpose of the survey.

The value proposition of advanced display technologies is not always clear.  While an argument based on accelerating the time to reach a given scientific outcome is compelling, it is rarely enough to justify the expense on its own.  However, it is becoming more relevant to respond rapidly to the influx of new data and science where the data needs to be analysed quickly, to ensure best-use of limited resources.  In this work, we have examined improvements to a new program aimed at detecting fast transients in real time requiring coordination of multiple observatories and astronomers and necessary rapid data analysis.    In this context, it was imperative that the {\em Deeper, Wider, Faster} team was able to make rapid determinations of likely candidates to trigger multi-wavelength imaging and spectroscopic follow-up observations.

\subsection{Potential improvements}
\label{lbl:obsimprovements}
The participating observers responded overwhelmingly positively to the combination of the TDW for O1, the independent workstations for O2, and the curved projection screen in both runs. However, there remain a number of opportunities for improving the display ecology through alternative choices of hardware and software.

A significant improvement would be to eliminate the bezels from the TDW in order to make it easier for the observers to see each entire, unbroken image at full resolution. This could be achieved by using ultra-thin bezel displays (though thin image breaks will still appear) or by using displays that match or exceed the resolution of the images being displayed. 

Currently the closest match to the DECam CCD image size is the 4K standard. Fully compliant  4K screens have a resolution of 4096 $\times$ 2160 pixels, which exceeds the resolution required to display the individual CCD images. However, these displays and projectors can be expensive. A more viable option would be the consumer version of 4K, more commonly called {\em ultra HD}. With a resolution of 3840 $\times$ 2160, these displays are not only more cost effective, but are also very close to the required resolution. In fact the images could be displayed at 96\% of full resolution, which should not result in too much degradation due to a small amount of pixel subsampling. 

The TDW offered a great deal of promise for the {\em Deeper, Wider, Faster} project.  It addressed the need to be able to display multiple high-resolution images for a short time and then refresh these with new images at a fast pace.  It allowed several researchers to search the images in parallel.  

However, during O1, software limitations of the TDW were apparent that made it unsuitable for viewing some of the astronomical data needed for {\em Deeper, Wider, Faster}. In particular, the combination of SAOImage DS9\footnote{\url{http://ds9.si.edu/site/Home.html}} and IRAF\footnote{\url{http://iraf.noao.edu/}} was critical to evaluating the software-detected potential candidates, but the TDW software did not provide an adequately performant mechanism to display this content.

There are several alternative software solutions for operating a TDW that were not explored during this campaign.  SAGE2 was chosen for the TDW after earlier testing had shown it to be the most suitable for general applications.  \citet{meade2014} discussed solutions such as CGLX\footnote{\url{http://vis.ucsd.edu/~cglx/}} and COVISE\footnote{\url{https://www.hlrs.de/en/covise/}} and their relative shortcomings.  Other solutions such as VisTrails\footnote{\url{https://www.vistrails.org/index.php/Main_Page}} were not tested due to the time constraints of the campaign, but would be worth investigating in the future.

However, using a TDW as a fully integrated display was ultimately not the most appropriate use of the infrastructure, as the refinement of the workflow revealed.  The improved display ecology for O2 highlighted the value of combining laptops with the new display configuration.  Therefore further investigations of alternative TDW software would have been fruitless.

\citet{bertin2015} discusses alternative methods of dealing with the presentation of large astronomical imagery, which aims to solve the problem of performance of presenting extremely large, remotely stored astronomical images.  In the context of {\em Deeper, Wider, Faster}, this approach might have rendered the transfer of the highly cadenced 4k images unnecessary for the purposes of review, provided a highly stable connection between Chile and Australia could be ensured. However, the {\em Mary} pipeline running on Swinburne's G2 cluster would still have required the transfer, and the CCD subtraction images were produced by this pipeline, making the transfer unavoidable.  Still aspects of this approach are being considered for future runs.

\subsection{Other applications}
\label{lbl:contexts}
After using the environment extensively, several potential astronomical applications for the use of the advanced display environment were identified.  These generally include any scenario where very small specific details contained within a very large context are critical to understanding the phenomenon being observed.  Examples include:

\begin{itemize}
\item Comparing absorption features in different transitions in quasar absorption spectra;
\item Large galaxy surveys looking for trends in shape and rotation curves; and
\item Viewing a large number of raw or reduced spectra from multi-object spectrographs
to identify unusual objects, place preliminary redshifts and run redshifting software; and
\item The TDW could help in the creation of training sets for machine learning software. Viewing thousands of images of real and non-real transient candidates in subtracted images to manually classify them for machine learning training sets would help produce more efficient automated software detections.
\end{itemize}

The successful use of a TDW as part of a collaborative workspace was consistent with  the findings of \citet{meade2014}: physical movement of the eyes, head and/or whole body was deemed preferable to using a keyboard and/or mouse to pan and zoom.  
There are several benefits to this approach:
\begin{enumerate}
\item It is easier to remember areas of the image already searched;
\item It is easier to maintain a sense of scale of objects being considered as the image scale is consistent and persistent;
\item Physical navigation is often quicker than virtual navigation; improving the time to analyse data; and
\item The activity is more stimulating than sitting and viewing in the one direction for prolonged periods; providing both physical relief and exercise.
\end{enumerate}

\section{Conclusion}
\label{sct:conclusion}


Establishing a workflow that employs a suitable display ecology combining advanced displays with standard displays has proven essential in advancing the science outcomes of the {\em Deeper, Wider, Faster} campaign.  The advantage of fast cadenced images can quickly become a disadvantage when manual inspection of the individual CCD images is required. The sheer volume of digital information makes it a challenging and cumbersome task for astronomers to achieve using standard, desktop-bound display technologies.  We developed a suitable display ecology for postage stamp and CCD image review, and it is clear that without this approach, such a demanding workflow would have been cumbersome and unlikely to have resulted in two successful campaigns.




Dedicated advanced displays, such as a TDW or large-area projection screen, may only solve one part of the image inspection problem. For the {\em Deeper, Wider, Faster} program, one display was more appropriate for parallel inspection of the multiple CCD images, while the other was more suited to displaying the numerous postage stamp candidates generated from the {\em Mary} automated source-detection pipeline.  However, it was discovered that when used in conjunction with the online spreadsheet logging tool, independent workstations with sufficient resolution for the CCD image review task was a better option. The use of standard laptops were well suited to interacting with the online spreadsheet. No display was well suited to all tasks and therefore only provided their maximum benefit when used in concert. The most appropriate devices are employed in an efficient manner to make all relevant information available in the most digestible, and actionable, form possible.

When it comes to processing vast amounts of data in useful timeframes, automation has allowed astronomy to advance well beyond human limitations.  Despite this, it remains the purview of the astronomer to determine the nature and direction of these advances.  Human inspection helps to train the software for better automated results and to place the detections in the larger context. Here, the eyes and experiences of astronomers remains a critical part of the discovery process.  Employing the right technology to enhance this capability is every bit as important as deploying more advanced telescopes.

\section*{Acknowledgments} 
The authors thank Nino Colella, Carlo Sgro and the Learning Environments team (University of Melbourne) for their support in using the curved projection screen, Ken Hodgson and Tony Mazzei (University of Melbourne) for assistance with building access and security, and Luc Renambot (University of Chicago) and Dr Ian Peake (RMIT) for their technical advice in the use of the SAGE2 environment. We also thank Dr Steven Manos (Director, Research Platform Services at the University of Melbourne) for the use of the OzIPortal Tiled Display Wall and the curved projection screen for the observing campaign. Research support to IA is provided by the Australian Astronomical Observatory (AAO).  This research was supported by use of the NeCTAR Research Cloud and by the Melbourne Node at the University of Melbourne.  The NeCTAR Research Cloud is a collaborative Australian research platform supported by the National Collaborative Research Infrastructure Strategy.

{}

\end{document}